\documentclass[eqsecnum,preprint,prd,aps,nofootinbib]{revtex4}
\usepackage{amsmath}
\usepackage{psfig}

\def\cstok#1{\leavevmode\thinspace\hbox{\vrule\vtop{\vbox{\hrule\kern1pt
\hbox{\vphantom{\tt/}\thinspace{\tt#1}\thinspace}}
\kern1pt\hrule}\vrule}\thinspace}

\begin{document}
\begin{center}
\bibliographystyle{article}

{\Large \textsc{The scalar wave equation in a non-commutative 
spherically symmetric space-time}}

\end{center}
\vspace{0.4cm}


\date{\today}

\author{Elisabetta Di Grezia,$^{1,2}$
\thanks{Electronic address: digrezia@na.infn.it}
Giampiero Esposito,$^{3,4}$ \thanks{
Electronic address: giampiero.esposito@na.infn.it}
Gennaro Miele,$^{4,3}$
\thanks{Electronic address: gennaro.miele@na.infn.it}}

\affiliation{${\ }^{1}$Universit\`a Statale di Bergamo, 
Facolt\`a di Ingegneria, Viale Marconi 5, 24044 
Dalmine (Bergamo)\\
${\ }^{2}$Istituto Nazionale di Fisica Nucleare, Sezione di Milano,
Via Celoria 16, 20133 Milano\\
${\ }^{3}$Istituto Nazionale di Fisica Nucleare,
Sezione di Napoli,\\
Complesso Universitario di Monte S. Angelo, Via Cintia, Edificio 6, 80126
Napoli, Italy\\
${\ }^{4}$Dipartimento di Scienze Fisiche, Complesso Universitario di Monte
S. Angelo,\\
Via Cintia, Edificio 6, 80126 Napoli, Italy}

\begin{abstract}
Recent work in the literature has studied a version of non-commutative 
Schwarzschild black holes where the effects of non-commutativity are described
by a mass function depending on both the radial variable $r$ and a 
non-commutativity parameter $\theta$. The present paper studies the
asymptotic behaviour of solutions of the zero-rest-mass scalar wave 
equation in such a modified Schwarzschild space-time in a neighbourhood 
of spatial infinity. The analysis is eventually reduced to finding 
solutions of an inhomogeneous Euler--Poisson--Darboux equation, where
the parameter $\theta$ affects explicitly the functional form of the 
source term. Interestingly, for finite values of $\theta$, there is full
qualitative agreement with general relativity: the conformal singularity
at spacelike infinity reduces in a considerable way the differentiability
class of scalar fields at future null infinity. In the physical space-time,
this means that the scalar field has an asymptotic behaviour with a  
fall-off going on rather more slowly than in flat space-time.
\end{abstract}

\maketitle
\bigskip
\vspace{2cm}

\section{Introduction}

The present paper, devoted to the scalar wave equation in a
non-commutative Schwarzschild space-time, is strongly motivated by three
branches of modern gravitational physics:
\vskip 0.3cm
\noindent
(i) In their investigation of quantum amplitudes in black-hole
evaporation \cite{Hawk05}, the authors of \cite{Farl04,Farl05} have
considered emission of scalar radiation in a black-hole collapse problem,
assuming non-spherical perturbations of the scalar field $\phi$ on the
final surface $\Sigma_{F}$, and that the intrinsic three-metric describes
an exactly spherically-symmetric spatial gravitational field.
\vskip 0.3cm
\noindent
(ii) In general relativity, unexpected features of the asymptotic structure
are already found to occur: massless scalar fields which have a Bondi-type
expansion in powers of $r^{-1}$ near null past infinity do not have such
an expansion near future null infinity; solutions which have physically 
reasonable Cauchy data may fail to have Bondi-type expansions near 
null infinity \cite{Stew79}. 
\vskip 0.3cm
\noindent 
(iii) According to the models studied in \cite{Nico06}, \cite{Smai04},
\cite{Spal06}, the non-commutativity of spacetime can be encoded in
the commutator of operators corresponding to spacetime coordinates,
i.e. (the integer $D$ below being even)
\begin{equation}
[x^{\mu},x^{\nu}]={\rm i} \; \theta^{\mu \nu}, \; 
\mu,\nu=1,2,...,D
\label{(1.1)}
\end{equation}
when the antisymmetric matrix $\theta^{\mu \nu}$ is taken to have a
block-diagonal form 
$$
\theta^{\mu \nu}={\rm diag}\Bigr(\theta_{1},...,\theta_{D/2}\Bigr)
$$
with
\begin{equation}
\theta_{i}=\theta 
\left(\begin{array}{cc}
0 & 1 \\
-1 & 0 \\
\end{array} \right)
\; \; \forall i=1,2,...,D/2,
\label{(1.2)}
\end{equation}
the parameter $\theta$ having dimension of length squared and being 
constant. As shown in \cite{Smai04}, the constancy of $\theta$ is very
important to obtain a consistent treatment of Lorentz invariance and
unitarity. The authors of \cite{Nico06} solve the Einstein equations
with mass density of a static, spherically symmetric, smeared
particle-like gravitational source as (hereafter, in agreement with our
earlier work \cite{Digr06}, we use $G=c={\hbar}=1$ units)
\begin{equation}
\rho_{\theta}(r)={M \over (4\pi \theta)^{3\over 2}}
{\rm e}^{-{r^{2}\over 4\theta}},
\label{(1.3)}
\end{equation}
which therefore plays the role of matter source. Their resulting 
spherically symmetric metric is
\begin{equation}
ds^{2}=-\left(1-{2m(r,\theta)\over r}\right)dt^{2}
+\left(1-{2m(r,\theta)\over r}\right)^{-1}dr^{2}
+r^{2}(d\Theta^{2}+\sin^{2}\Theta d\varphi^{2}),
\label{(1.4)}
\end{equation}
where, in terms of the lower incomplete gamma function
\begin{equation}
\gamma \left({3\over 2},{r^{2}\over 4\theta}\right) \equiv
\int_{0}^{{r^{2}\over 4\theta}}\sqrt{t} {\rm e}^{-t} \; dt,
\label{(1.5)}
\end{equation}
we define the mass function \cite{Nico06, Digr06}
\begin{equation}
m(r,\theta) \equiv {2M \over \sqrt{\pi}}
\gamma \left({3\over 2},{r^{2}\over 4\theta}\right).
\label{(1.6)}
\end{equation}

Thus, if one tries to study emission of scalar radiation as in
\cite{Farl04, Farl05} but in the presence of a non-vanishing $\theta$
parameter (cf \cite{Digr06} for the pure gravity case), one is naturally
led to study a scalar wave equation in a spherically symmetric spacetime
whose metric is affected by $\theta$, which is the goal of the present
paper. Section 2 builds conformal infinity for the space-time with
metric (1.4). Section 3, following \cite{Stew79}, 
turns the scalar wave equation into an
inhomogeneous Euler--Poisson--Darboux equation. Section 4 solves such
an equation and shows under which conditions there is full qualitative
agreement with general relativity. Concluding 
remarks and open problems are presented in section 5, while the appendices
describe relevant mathematical details.

\section{Conformal infinity}

Inspired by general relativity, we define a new radial coordinate 
$r^{*}$ in such a way that
\begin{equation}
dr^{*}={dr \over 1-{2m(r,\theta)\over r}}.
\label{(2.1)}
\end{equation}
This equation is solved by 
\begin{equation}
r^{*}=r+2 \int {m(r,\theta)\over r-2m(r,\theta)}dr ,
\label{(2.2)}
\end{equation}
and the metric (1.4) can be re-expressed in the form
\begin{equation}
ds^{2}=-\left(1-{2m(r,\theta)\over r}\right)du dv
+r^{2}(d\Theta^{2}+\sin^{2}\Theta d\varphi^{2}),
\label{(2.3)}
\end{equation}
where $\theta \in [0,\pi], \varphi \in [0,2\pi]$, and we have defined
the `retarded' coordinate
\begin{equation}
u \equiv t-r^{*} \; \; \in ]-\infty,\infty[ ,
\label{(2.4)}
\end{equation}
and the `advanced' coordinate
\begin{equation}
v \equiv t+r^{*} \; \; \in ]-\infty,\infty[.
\label{(2.5)}
\end{equation}
The equations (2.2), (2.4) and (2.5) yield
\begin{equation}
{2\over (v-u)}={1\over r \left[1+{2\over r} 
\int {m(r,\theta)\over r-2m(r,\theta)}dr \right]}.
\label {(2.6)}
\end{equation}
To lowest order, Eq. (2.6) is solved by ${1\over r} \approx 
{2\over (v-u)}$, and on defining 
\begin{equation}
F(u,v,\theta) \equiv \left . \int {m(r,\theta) \over r-2m (r,\theta)}dr
\right |_{r={(v-u)\over 2}} ,
\label{(2.7)}
\end{equation}
one finds, by iterated approximations, the asymptotic expansion
\begin{equation}
{1\over r} \sim {2\over (v-u)}+8{F(u,v,\theta)\over (v-u)^{2}}
+{\rm O} \left({F^{2}(u,v,\theta)\over (v-u)^{3}}\right).
\label{(2.8)}
\end{equation}

The limit $v \rightarrow + \infty$ with $u,\Theta,\varphi$ fixed defines 
future null infinity ${\cal I}^{+}$; the limit $u \rightarrow + \infty$
with $v,\Theta,\varphi$ fixed defines past null infinity 
${\cal I}^{-}$, while the limit $u \rightarrow -\infty, v \rightarrow
+ \infty$ with $(u+v),\Theta,\varphi$ fixed defines 
spacelike infinity, i.e. the point $I^{0}$. 
The figures below show the behaviour of the denominator 
$y \equiv 1-{2m(r,\theta)\over r}$ in Eq. (2.1) for various values 
of $\theta$. Interestingly, the occurrence of $\theta$ does not introduce
new singularities with respect to general relativity.

It is actually simpler to introduce the coordinates $u$ and $v$ 
separately, which yields the conformally rescaled, 
``unphysical'' metrics (here
$f \equiv r^{-1}$, and $d\Sigma^{2}$ is the metric on a unit
two-sphere)
\begin{equation}
d{\widetilde s}^{2}= f^{2}\left[-(1-2mf)du^{2}-2du \; dr
+r^{2} d\Sigma^{2}\right]
=-(f^{2}-2mf^{3})du^{2}+2du \; df +d\Sigma^{2},
\label{(2.9)}
\end{equation}
and
\begin{equation}
d{\widetilde S}^{2}=-(f^{2}-2mf^{3})dv^{2}-2dv \; df 
+d\Sigma^{2}.
\label{(2.10)}
\end{equation}
These metrics are manifestly regular and analytic on their respective 
hypersurfaces $f=0$, since their determinants are equal to 
$- \sin^{2}\Theta$ for all $f$, including 
$f=0$. The physical space-time
corresponds to $f>0$ in (2.9), and we can extend the manifold to include
${\cal I}^{+}$, given when $f=0$. Similarly, in (2.10), the physical
space-time corresponds to $f>0$ and can 
be extended to include ${\cal I}^{-}$,
given when $f=0$. Only the boundary ${\cal I} 
\equiv {\cal I}^{+} \cup {\cal I}^{-}$ is adjoined to the space-time. 

In common with general relativity, we note here a difficulty that is 
encountered if we try to identify ${\cal I}^{-}$ with ${\cal I}^{+}$.
If we do extend the region of definition of (2.9) to include negative
values of $f$, and then make the replacement 
$f \rightarrow -f$, we see
that the metric has the form (2.10) (with $u$ in place of $v$) but with
a mass function $-m$ in place of $m$. Thus, the extension across ${\cal I}$
involves a reversal of the sign of the mass function, which is incompatible
with Eq. (1.6) unless we advocate a discontinuity in the derivative of the
curvature \cite{Penr86} 
across $\cal I$. It is therefore not reasonable to identify
${\cal I}^{+}$ with ${\cal I}^{-}$.

To sum up, we have two disjoint boundary hypersurfaces ${\cal I}^{+}$
and ${\cal I}^{-}$, each of which is a cylinder with topology
$S^{2} \times {\bf R}$. It is clear from (2.9) and (2.10) that each of
${\cal I}^{\pm}$ is a null hypersurface (the induced metric at $f=0$
being degenerate). These null hypersurfaces are generated by rays
(given by $\Theta,\phi ={\rm constant}$, $f=0$) whose tangents are 
normals to the hypersurfaces. These rays may be taken to be the 
${\bf R}$'s of the topological product $S^{2} \times {\bf R}$.

\begin{figure}[!h]
\centerline{\hbox{\psfig{figure=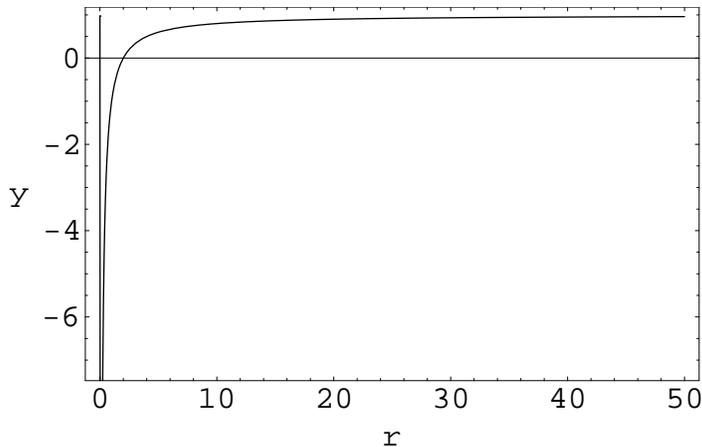,width=0.6\textwidth}}}
\caption{Plot of the denominator $y \equiv 1-{2m(r,\theta)\over r}$ in
Eq. (2.1) when $\theta=10^{-7}$.}
\end{figure}

\begin{figure}[!h]
\centerline{\hbox{\psfig{figure=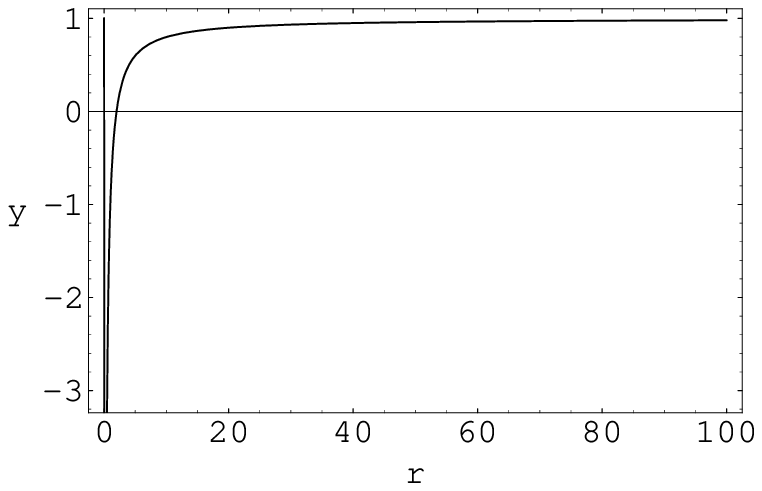,width=0.6\textwidth}}}
\caption{Plot of the denominator $y \equiv 1-{2m(r,\theta)\over r}$ in
Eq. (2.1) when $\theta=10^{-4}$.}
\end{figure}

\begin{figure}[!h]
\centerline{\hbox{\psfig{figure=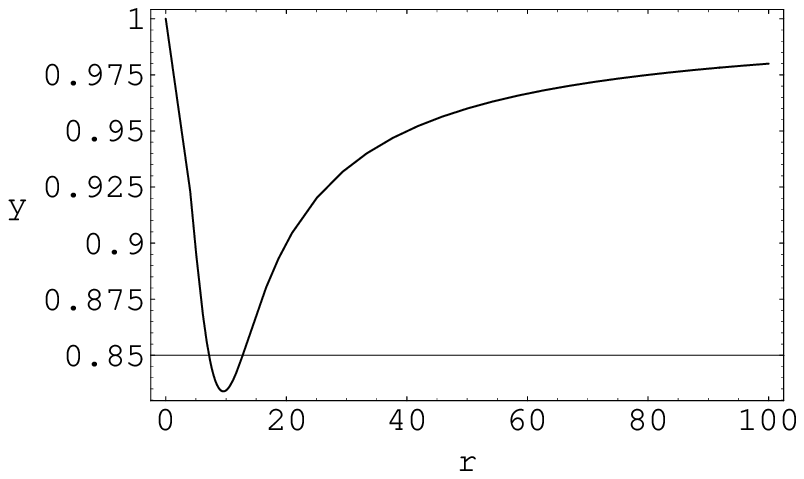,width=0.6\textwidth}}}
\caption{Plot of the denominator $y \equiv 1-{2m(r,\theta)\over r}$ in
Eq. (2.1) when $\theta=10$.}
\end{figure}

\begin{figure}[!h]
\centerline{\hbox{\psfig{figure=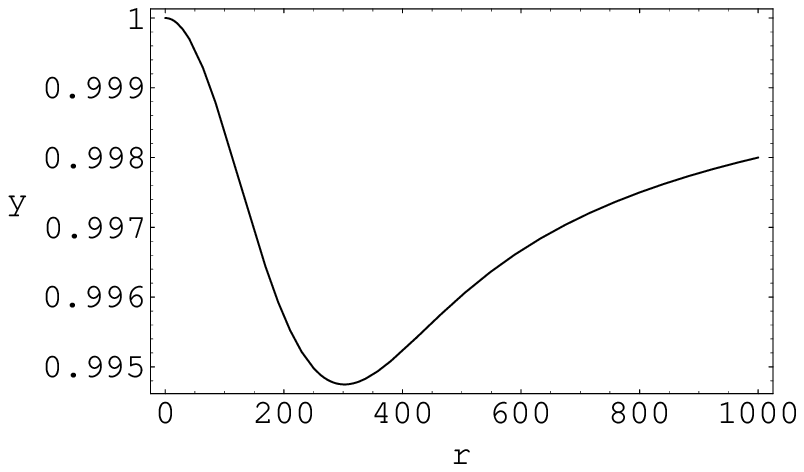,width=0.6\textwidth}}}
\caption{Plot of the denominator $y \equiv 1-{2m(r,\theta)\over r}$ in
Eq. (2.1) when $\theta=10^{4}$.}
\end{figure}

\clearpage

\section{Inhomogeneous Euler--Poisson--Darboux equation}

The coordinates $(u,v)$ defined in (2.4) and (2.5) are 
not the most convenient for
discussing the limits which define conformal infinity \cite{Stew79}. We
therefore define (cf. \cite{Stew79}) a function $w_{\theta}(x)$ by 
requiring that $w_{\theta}(x=r^{-1})$ should be equal to 
$r^{*}$ in (2.2), i.e.
\begin{equation}
w_{\theta}(x) \equiv  \int {dx \over 
x^{2}(2xm(x^{-1},\theta)-1)} ,
\label{(3.1)}
\end{equation}
which implies
\begin{equation}
g_{\theta}(x) \equiv -w_{\theta}'(x) ={1\over x^{2}(1-2xm)}.
\label{(3.2)}
\end{equation}
Equation (3.1) defines a one-parameter family of monotone decreasing
$C^{\infty}$ functions taking values over the whole real line. The
monotone decreasing character of $w_{\theta}$ is proved by imposing that
$1-2xm >0$. This is indeed satisfied for sufficiently small values of
$\theta$, so that $1-2xm \approx 1-2xM$, which is positive
provided $x < {1\over 2M}$.
A $C^{\infty}$ inverse function therefore exists, which makes it possible
to define new coordinates $a,b$ according to (cf \cite{Stew79})
\begin{equation}
w_{\theta}(x=a) \equiv 
\left . \int {dx \over x^{2}(2xm-1)} \right |_{x=a}
={v \over 2}={t\over 2}+{r^{*}\over 2},
\label{(3.3)}
\end{equation}
\begin{equation}
w_{\theta}(x=b) \equiv 
\left . \int {dx \over x^{2}(2xm-1)} \right |_{x=b}
=-{u \over 2}=-{t\over 2}+{r^{*}\over 2},
\label{(3.4)}
\end{equation}
where the integrals (3.3) and (3.4) involve the mass function 
$m=m(r=x^{-1})$. On defining $f \equiv r^{-1}$
as in section 2, one finds from 
(2.2), (3.1), (3.3) and (3.4) that
\begin{equation}
w_{\theta}(f(a,b))=r^{*}(a,b)=w_{\theta}(x=a)+w_{\theta}(x=b).
\label{(3.5)}
\end{equation}
Moreover, from (3.2)--(3.4), the metric (2.3) in the $(u,v,\Theta,\varphi)$
coordinates takes the following form in the $(a,b,\Theta,\varphi)$
coordinates:
\begin{equation}
ds^{2}=4(1-2mf)g_{1}(a)g_{2}(b)da \; db
+f^{-2}d\Sigma^{2},
\label{(3.6)}
\end{equation} 
having defined
\begin{equation}
M_{1}(a) \equiv m(a^{-1},\theta), \;
g_{1}(a) \equiv a^{-2}(1-2aM_{1}(a))^{-1}=g_{\theta}(a),
\label{(3.7)}
\end{equation}
\begin{equation}
M_{2}(b) \equiv m(b^{-1},\theta), \;
g_{2}(b) \equiv b^{-2}(1-2bM_{2}(b))^{-1}=g_{\theta}(b).
\label{(3.8)}
\end{equation}

In the analysis of the scalar wave equation $\cstok{\ } \psi=0$, 
we now rescale the scalar field $\psi$ according to
\begin{equation}
{\widetilde \psi}=\Omega^{-1}\psi,
\label{(3.9)}
\end{equation}
where $\Omega$ is a real positive function such that
\begin{equation}
\Omega=0, \; \Omega_{,k} \not = 0, \; g^{ik}\Omega_{,i} \Omega_{,k}=0
\; {\rm on} \; {\cal I}^{\pm}.
\label{(3.10)}
\end{equation}
The `unphysical' scalar field $\widetilde \psi$ satisfies the conformally
invariant wave equation in $4$ spacetime dimensions, i.e.
\begin{equation}
\left(\cstok{\ } +{R\over 6} \right){\widetilde \psi}=0,
\label{(3.11)}
\end{equation}
where $\cstok{\ }$ is the D'Alembert wave operator \cite{Stew79}, 
and $R$ is the scalar curvature in the `unphysical' space-time
with line element
\begin{equation}
d{\widetilde s}^{2}=\Omega^{2}ds^{2}=4 \Omega^{2}(1-2mf)
g_{1}(a)g_{2}(b)da \; db
+\Omega^{2}f^{-2}d\Sigma^{2}.
\label{(3.12)}
\end{equation}
On choosing the conformal factor in the form $\Omega=(a+b)f$, we 
therefore obtain the metric tensor
\begin{equation}
g_{\mu \nu}=
\begin{pmatrix}
0 \hfill & G(a,b) \hfill & 0 \hfill & 0 \hfill \\ 
G(a,b) \hfill & 0 \hfill & 0 \hfill & 0 \hfill \\
0 \hfill & 0 \hfill & (a+b)^{2} \hfill & 0 \hfill \\
0 \hfill & 0 \hfill & 0 \hfill & (a+b)^{2}\sin^{2} \Theta 
\end{pmatrix}
\label{(3.13)}
\end{equation}
having introduced
\begin{equation}
m(a,b) \equiv {2M\over \sqrt{\pi}}
\gamma \left({3\over 2},{1\over 4 \theta f^{2}(a,b)}\right),
\label{(3.14)}
\end{equation}
\begin{equation}
F(a,b) \equiv f^{2}(a,b)-2m(a,b)f^{3}(a,b),
\label{(3.15)}
\end{equation}
\begin{equation}
G(a,b) \equiv 2(a+b)^{2} g_{1}(a)g_{2}(b)F(a,b).
\label{(3.16)}
\end{equation}

By virtue of spherical symmetry, we look for solutions of Eq. (3.11) as a
linear combination of factorized terms as 
\begin{equation}
{\widetilde \psi}_{\theta}(a,b,\Theta,\varphi)
={\chi_{\theta}(a,b)\over (a+b)}Y_{lm}(\Theta,\varphi),
\label{(3.17)}
\end{equation}
where $Y_{lm}(\Theta,\varphi)$ are the spherical harmonics
on $S^{2}$. Substitution of the ansatz (3.17) into Eq. (3.11) gives
\begin{equation}
L[\chi]=S_{\theta}(a,b)\chi,
\label{(3.18)}
\end{equation}
where $L$ is the Euler--Poisson--Darboux operator \cite{Stew79}
\begin{equation}
L \equiv {\partial^{2}\over \partial a \partial b}
-{l(l+1)\over (a+b)^{2}},
\label{(3.19)}
\end{equation}
which depends implicitly on $\theta$ through $a,b$ defined in (3.3), (3.4),
while $S_{\theta}$ is the $\theta$-dependent source term
\begin{eqnarray}
S_{\theta}(a,b) & \equiv & l(l+1){\left({G\over 2}-1 \right)
\over (a+b)^{2}} +{1\over 12}G R(\theta) \nonumber \\
&=& l(l+1)\left[g_{1}(a)g_{2}(b)F(a,b)-{1\over (a+b)^{2}} \right]
\nonumber \\
&+& {1\over 6}(a+b)^{2} g_{1}(a)g_{2}(b)F(a,b)R(\theta),
\label{(3.20)}
\end{eqnarray}
having denoted by $R(\theta)$ the scalar curvature in Eq. (B24). 
Inspired by \cite{Stew79}, we now write the solution of Eq. (3.18)
as the sum $\chi^{0}+L^{-1}S$, where $\chi^{0}$ is the general solution
of the homogeneous Euler--Poisson--Darboux equation $L[\chi]=0$, while
$L^{-1}$ is an integral operator with kernel given by the 
Riemann--Green function (see appendix) of $L$ \cite{Cour61}:
\begin{equation}
\chi_{\theta}(a,b)=\chi^{0}(a,b)- \int \int_{D(a,b)} R(a,b;a',b') 
S_{\theta}(a',b')\chi_{\theta}(a',b')da' \; db',
\label{(3.21)}
\end{equation}
having defined
\begin{equation}
D(a,b) \equiv \{ a',b': 0 \leq a \leq a' \leq b' \leq b \}.
\label{(3.22)}
\end{equation}
As is described in \cite{Stew79}, $\chi^{0}(a,b)$ has the general form
\begin{equation}
\chi^{0}(a,b)=(a+b)^{l+1}\left[\left({\partial \over \partial a}\right)^{l}
{A(a)\over (a+b)^{l+1}}+\left({\partial \over \partial b}\right)^{l}
{B(b)\over (a+b)^{l+1}}\right],
\label{(3.23)}
\end{equation}
with $A$ and $B$ arbitrary $C^{l+1}$ functions. Moreover, the 
Riemann--Green function of the operator $L$ defined in (3.19) is obtained 
from the Legendre polynomial of degree $l$ according to \cite{Cops58}
\begin{equation}
R(a,b;a',b')=P_{l}(z(a,b;a',b')),
\label{(3.24)}
\end{equation}
having defined \cite{Stew79}
\begin{equation}
z(a,b;a',b') \equiv {(a-a')(b-b')+(a+b')(a'+b) \over (a+b)(a'+b')}.
\label{(3.25)}
\end{equation}

\section{Qualitative analysis of the $l=0$ solution}

Hereafter we consider for simplicity the case $l=0$; the comparison with
our Ref. \cite{Stew79} is then easier, and all main features are already
displayed. Strictly, we consider an asymptotic characteristic initial-value
problem where data are specified on past null infinity for 
$a \in [0,a_{0}]$ and on the outgoing null hypersurface 
$a=a_{0}={\rm constant}$. If $l=0$, it is clear from (3.23) that the
characteristic data can be set to $1$: $\chi^{0}(a,b)=1$, while the
Riemann--Green function in (3.24) reduces to $1$ \cite{Stew79}:
\begin{equation}
R_{l=0}(a,b;a',b')=P_{0}(z(a,b;a',b'))=1.
\label{(4.1)}
\end{equation}
The inhomogeneous wave equation (3.18) with $l=0$ can be solved with the
help of a contraction mapping, i.e. \cite{Stew79}
\begin{equation}
\chi(a,b)=\chi^{0}(a,b)+\sum_{n=1}^{\infty}\chi^{n}(a,b)
=1+\sum_{n=1}^{\infty}\chi^{n}(a,b),
\label{(4.2)}
\end{equation}
where
\begin{equation}
\chi^{n}(a,b)=\int \int_{D(a,b)}S_{\theta}(a',b')\chi^{n-1}(a',b')
da' \; db' = {\rm O}((a+b)^{n}),
\label{(4.3)}
\end{equation}
and the series in (4.2) is known to be uniformly convergent near 
spacelike infinity in general relativity \cite{Stew79}. Moreover,
in general relativity the partial derivative $\chi_{,a}$ as
$a \rightarrow 0$ and $b$ is fixed remains bounded, as well as the
partial derivative $\chi_{,b}$ as $b \rightarrow 0$ and $a$ is fixed.
The second derivative $\chi_{,aa}$, however, diverges near future 
null infinity, which implies that the presence of a conformal singularity
at spacelike infinity affects the behaviour of scalar fields on future
null infinity, reducing by a considerable amount their differentiability
class. Such a property is proved by exploiting the integral 
representation of $\chi_{,aa}$, i.e.
\begin{equation}
\chi_{,aa}=- \int_{a}^{b}\Bigr[S_{,a}(a,b')\chi(a,b')
+S(a,b')\chi_{,a}(a,b')\Bigr]db'.
\label{(4.4)}
\end{equation}
By insertion of (4.2) into (4.4), and bearing in mind (4.3), one finds 
that the possible singularities of $\chi_{,aa}$ are ruled by the 
integrals \cite{Stew79}
\begin{equation}
I_{0}(a,b) \equiv \int_{0}^{b}S_{,a}(a,b')db',
\label{(4.5)}
\end{equation}
\begin{equation}
I_{1}(a,b) \equiv \int_{0}^{b}(a+b')S_{,a}(a,b')db',
\label{(4.6)}
\end{equation}
where $I_{0}$ remains finite as $a \rightarrow 0$, whereas $I_{1}$
displays a logarithmic singularity as $a \rightarrow 0$.

If $l=0$ in Sec. 3 we find, for finite values of $\theta$, the
counterpart of (4.5) given by the integral
\begin{equation}
{\widetilde I}_{0}(a,b) \equiv \int_{0}^{b}S_{\theta,a}(a,b')db',
\label{(4.7)}
\end{equation}
where, from the asymptotic formulae as $a \rightarrow 0$ and $b$ is
fixed, we find, for all finite values of $\theta$,
\begin{equation}
f(a,b) \sim {ab \over (a+b)},
\label{(4.8)}
\end{equation}
\begin{equation}
m(a,b) \sim {2M \over \sqrt{\pi}}
\int_{0}^{{(a+b)^{2}\over 4 \theta a^{2}b^{2}}} \sqrt{t}
{\rm e}^{-t} \; dt \sim M,
\label{(4.9)}
\end{equation}
\begin{equation}
F(a,b) \equiv (f^{2}-2m f^{3})(a,b) \sim
{a^{2}b^{2} \over (a+b)^{2}}(1+{\rm O}(a)),
\label{(4.10)}
\end{equation}
\begin{equation}
S_{\theta} \sim {2Mab \over (a+b)^{3}},
\label{(4.11)}
\end{equation}
\begin{equation}
S_{\theta,a} \sim 2Mb \left[-{2\over (a+b)^{3}}+{3b \over (a+b)^{4}}
\right],
\label{(4.12)}
\end{equation}
and hence
\begin{equation}
{\widetilde I}_{0} \sim -{2Mb^{2}\over (a+b)^{3}} 
\; {\rm as} \; a \rightarrow 0,
\label{(4.13)}
\end{equation}
in agreement with the analysis in \cite{Stew79} for general relativity.
These approximations should be
abandoned only if $\theta$ is so large that (cf. (4.9))
\begin{equation}
\lim_{a \to 0} \theta a^{2}={\rm constant}.
\label{(4.14)}
\end{equation}
Furthermore, the counterpart of (4.6) is given by the integral 
\begin{equation}
{\widetilde I}_{1}(a,b) \equiv \int_{0}^{b}
(a+b')S_{\theta,a}(a,b')db' \sim -2M \log(a) \; {\rm as}
\; a \rightarrow 0,
\label{(4.15)}
\end{equation} 
again in full agreement with \cite{Stew79}. Note that a more
accurate asymptotic expansion of the source term would be
\begin{equation}
S_{\theta} \sim {2Mab \over (a+b)^{3}}(1-2b M_{2}(b))^{-1},
\label{(4.16)}
\end{equation}
but this does not modify the leading terms as $a \rightarrow 0$
in (4.13) and (4.15).

The figures below show the behaviour of $\chi, \chi_{,a}, \chi_{,b}$
and $\chi_{,aa}$.

\begin{figure}[!h]
\centerline{\hbox{\psfig{figure=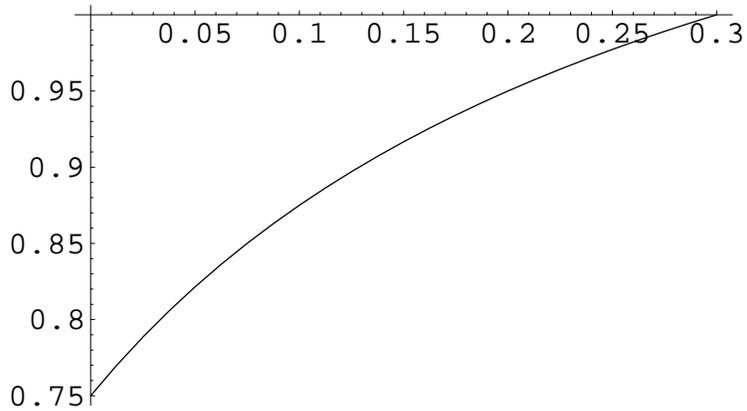,width=0.6\textwidth}}}
\caption{Plot of the solution $\chi(a,b)$ in (4.2) and (4.3) with 
$n=1$ at small $a$, when $\theta$ takes finite values and $b=0.3$.}
\end{figure}

\begin{figure}[!h]
\centerline{\hbox{\psfig{figure=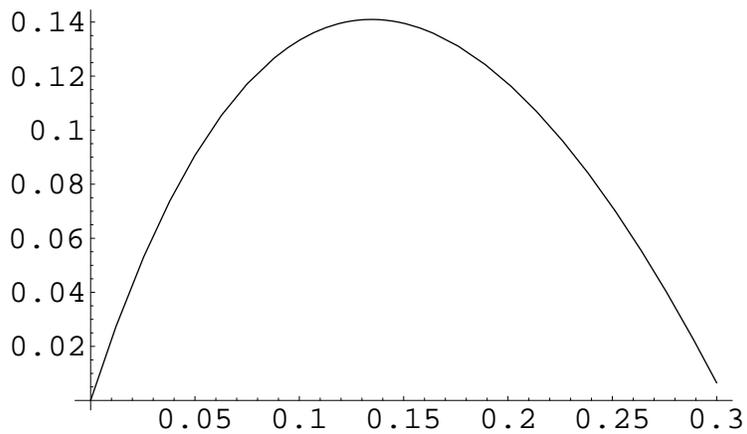,width=0.6\textwidth}}}
\caption{Plot of the partial derivative $\chi_{,a}$ when 
$\theta=10^{-3}$ and $b=0.3$. Such a derivative is clearly bounded,
according to the theoretical expectations \cite{Stew79}. The identical
behaviour is displayed by $\chi_{,b}$ when $\theta=10^{-3}$ and
$a=0.3$.}
\end{figure}
 
\begin{figure}[!h]
\centerline{\hbox{\psfig{figure=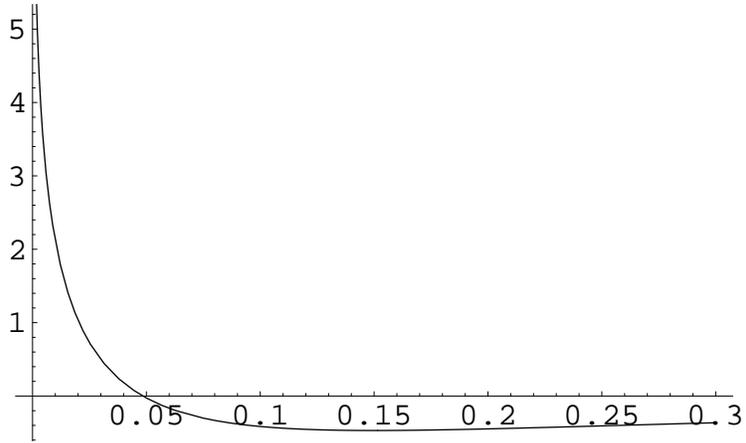,width=0.6\textwidth}}}
\caption{Plot of the partial derivative $\chi_{,aa}$ when 
$\theta=10^{-3}$ and $b=0.3$. The logarithmic singularity as
$a \rightarrow 0^{+}$ is clearly displayed, and it occurs at all 
finite values of $\theta$.}
\end{figure}

\clearpage

\section{Concluding remarks}
 
Ever since Penrose \cite{Penr64} developed a geometrical picture of
an isolated system in general relativity as a space-time admitting
future and past null infinity (with the associated fall-off of the
metric along null geodesics going off to infinity), there has been
always great interest in this coordinate-free way of bringing infinity
to a `finite distance' and discussing the asymptotic structure of
space-time. Moreover, the conceptual revolution brought about by 
non-commutative geometry \cite{Land97, Conn06, Lizz07} has led to an
assessment of the very concept of space-time manifold
\cite{Asch06}, with 
`corrections' to it evaluated, for example, along the lines of the 
work in Refs. \cite{Nico06, Smai04, Spal06}. 
Within this broad framework, the contributions of our paper are as follows.
\vskip 0.3cm
\noindent
(i) Construction of conformal infinity for the spherically symmetric
space-time which incorporates noncommutative-geometry corrections to
Schwarzschild space-time.
\vskip 0.3cm
\noindent
(ii) Evaluation of the source term (3.20) in the inhomogeneous
Euler--Poisson--Darboux equation which describes the scalar wave equation
in the unphysical space-time obtained after conformal rescaling of
the original metric (1.4).
\vskip 0.3cm
\noindent
(iii) Qualitative analysis of the asymptotic characteristic initial-value
problem in the $l=0$ case, finding again the logarithmic singularity as
shown in Eq. (4.15). In the original, `physical' space-time with metric
(1.4), such a singularity implies that the large-$r$ behaviour of the
scalar field is described by the asymptotic expansion \cite{Stew79}
\begin{equation}
\psi \sim {c_{1}\over r}+{c_{2}\over r^{2}}
+{d_{1}\log(r)\over r^{3}}+{\rm O}(r^{-3}),
\label{(5.1)}
\end{equation}
and therefore the field falls off at large $r$ rather more slowly than 
in flat space-time \cite{Stew79}. 
\vskip 0.3cm
\noindent
(iv) Numerical support for all results in sections 3 and 4, as shown
by figures 5--7 at the end of section 4.

Our results are thus an encouraging progress towards a rigorous theory
of wavelike phenomena in noncommutative geometry, along the lines of
the conformal-infinity program of Penrose for general 
relativity \cite{Penr86}. 
Hopefully, the physical applications to isolated gravitating systems 
in a noncommutative framework, and possibly to black-hole evaporation, 
will also become clear in the years to come.
 
\acknowledgments
The authors are grateful to the INFN for financial support.
The work of G. Miele has been partially supported by
PRIN {\it FISICA ASTROPARTICELLARE}.

\appendix
\section{The Riemann--Green function}

An hyperbolic equation in two independent variables can be always cast
in the canonical form \cite{Cour61}
\begin{equation}
L[\chi]=\left({\partial^{2}\over \partial x \partial y}
+a(x,y){\partial \over \partial x}+b(x,y){\partial \over \partial y}
+c(x,y) \right)\chi(x,y)=H(x,y).
\label{(A1)}
\end{equation}
One can then use the Riemann integral representation of the solution
\cite{Cour61}. For this purpose, denoting by $L^{\dagger}$ the adjoint
of the operator $L$ in (A1), which acts according to
\begin{equation}
L^{\dagger}[\chi]=\chi_{,xy}-(a \chi)_{,x} -(b \chi)_{,y}+c \chi,
\label{(A2)}
\end{equation}
one has to find the Riemann kernel $R(x,y;\xi,\eta)$ subject to the
following conditions ($(\xi,\eta)$ being the coordinates of a point $P$
such that the characteristics through it intersect a curve $C$ at points
$A$ and $B$, $AP$ being a segment with constant $y$, and $BP$ being a
segment with constant $x$):
\vskip 0.3cm
\noindent
(i) As a function of $x$ and $y$, $R$ satisfies the adjoint equation
\begin{equation}
L_{(x,y)}^{\dagger}[R]=0,
\label{(A3)}
\end{equation}
\vskip 0.3cm
\noindent
(ii) $R_{,x}=bR$ on $AP$, i.e.
\begin{equation}
R_{,x}(x,y;\xi,\eta)=b(x,\eta)R(x,y;\xi,\eta) \; {\rm on} \; 
y=\eta,
\label{(A4)}
\end{equation}
and $R_{,y}=aR$ on $BP$, i.e.
\begin{equation}
R_{,y}(x,y;\xi,\eta)=a(\xi,y)R(x,y;\xi,\eta) \; {\rm on} \;
x=\xi,
\label{(A5)}
\end{equation}
\vskip 0.3cm
\noindent
(iii) $R$ equals $1$ at $P$, i.e.
\begin{equation}
R(\xi,\eta;\xi,\eta)=1.
\label{(A6)}
\end{equation}
It is then possible to express the solution of Eq. (A1) in the form
\cite{Cour61}
\begin{eqnarray}
\chi(P)&=& {1\over 2}[\chi(A)R(A)+\chi(B)R(B)]
+ \int_{AB} \biggr( \left[{R\over 2}\chi_{,x}
+\left(bR -{1\over 2}R_{,x}\right)\chi \right]dx \nonumber \\
&-&  \left[{R\over 2}\chi_{,y}+\left(aR -{1\over 2}R_{,y}
\right)\chi \right]dy \biggr)
+ \int \int_{\Omega}R(x,y;\xi,\eta)H(x,y)dxdy,
\label{(A7)}
\end{eqnarray}
where $\Omega$ is a domain with boundary. 

Note that Eqs. (A4) and (A5) are ordinary differential equations for
the Riemann function $R(x,y;\xi,\eta)$ along the characteristics 
parallel to the coordinate axes. By virtue of (A6), their integration
yields
\begin{equation}
R(x,\eta;\xi,\eta)={\rm exp} \int_{\xi}^{x}b(\lambda,\eta)d\lambda,
\label{(A8)}
\end{equation}
\begin{equation}
R(\xi,y;\xi,\eta)={\rm exp} \int_{\eta}^{y} a(\lambda,\xi)d\lambda,
\label{(A9)}
\end{equation}
which are the values of the Riemann kernel $R$ along the characteristics
through $P$. Equation (A7) yields instead the solution of Eq. (A1) for
arbitrary initial values given along an arbitrary non-characteristic
curve $C$, by means of a solution $R$ of the adjoint equation (A3) which
depends on $x,y$ and two parameters $\xi,\eta$. Unlike $\chi$, the
Riemann function solves a characteristic initial-value problem
\cite{Cour61}.  

\section{Scalar curvature in the unphysical space-time}

In section 3, the evaluation of the scalar curvature $R=R(\theta)$ in the
unphysical space-time with metric (3.13) is as follows. 
With the notation in Eqs. (3.13)--(3.16), the non-vanishing connection
coefficients turn out to be (no summation over repeated indices)
\begin{equation}
\Gamma_{\; \Theta \Theta}^{a}=\Gamma_{\; \Theta \Theta}^{b}
=-(a+b)G^{-1}, \; 
\label{(B1)}
\end{equation}
\begin{equation}
\Gamma_{\; \varphi \varphi}^{a}=\Gamma_{\; \varphi \varphi}^{b}
=-(a+b)G^{-1}\sin^{2}\Theta,
\label{(B2)}
\end{equation}
\begin{equation}
\Gamma_{\; aa}^{a}=G^{-1}G_{,a}, \; 
\Gamma_{\; bb}^{b}=G^{-1}G_{,b},
\label{(B3)}
\end{equation}
\begin{equation}
\Gamma_{\; a \Theta}^{\Theta}=\Gamma_{\; \Theta a}^{\Theta}
=\Gamma_{\; b \Theta}^{\Theta}=\Gamma_{\; \Theta b}^{\Theta}
=\Gamma_{\; a \varphi}^{\varphi}=\Gamma_{\; \varphi a}^{\varphi}
=\Gamma_{\; b \varphi}^{\varphi}=\Gamma_{\; \varphi b}^{\varphi}
=(a+b)^{-1},
\label{(B4)}
\end{equation}
\begin{equation}
\Gamma_{\; \varphi \varphi}^{\Theta}=-\sin \Theta \cos \Theta,
\label{(B5)}
\end{equation}
\begin{equation}
\Gamma_{\; \Theta \varphi}^{\varphi}=\Gamma_{\; \varphi \Theta}^{\varphi}
=\cot \Theta.
\label{(B6)}
\end{equation}
The resulting Riemann tensor is evaluated from the general formula in a
coordinate basis
\begin{equation}
R_{\; \mu \nu \rho}^{\lambda}=\Gamma_{\; \mu \rho , \nu}^{\lambda}
-\Gamma_{\; \mu \nu , \rho}^{\lambda}
+\Gamma_{\; \mu \rho}^{\alpha} \Gamma_{\; \alpha \nu}^{\lambda}
-\Gamma_{\; \mu \nu}^{\alpha} \Gamma_{\; \alpha \rho}^{\lambda}.
\label{(B7)}
\end{equation}
Since we evaluate the scalar curvature $R=g^{\mu \nu}R_{\mu \nu}$, which
in our case is equal to (hereafter, no summation over repeated indices
$a$ or $b$ or $\Theta$ or $\varphi$)
\begin{equation}
R=2g^{ab}R_{ab}+g^{\Theta \Theta}R_{\Theta \Theta}
+g^{\varphi \varphi} R_{\varphi \varphi},
\label{(B8)}
\end{equation}
we only need the $12$ components of Riemann occurring in
\begin{equation}
R_{ab}=R_{\; aab}^{a}+R_{\; abb}^{b}+R_{\; a \Theta b}^{\Theta}
+R_{\; a \varphi b}^{\varphi},
\label{(B9)}
\end{equation}
\begin{equation}
R_{\Theta \Theta}=R_{\; \Theta a \Theta}^{a}
+R_{\; \Theta b \Theta}^{b}+R_{\; \Theta \Theta \Theta}^{\Theta}
+R_{\; \Theta \varphi \Theta}^{\varphi},
\label{(B10)}
\end{equation}
\begin{equation}
R_{\varphi \varphi}=R_{\; \varphi a \varphi}^{a}
+R_{\; \varphi b \varphi}^{b}+R_{\; \varphi \Theta \varphi}^{\Theta}
+R_{\; \varphi \varphi \varphi}^{\varphi}.
\label{(B11)}
\end{equation}
Among these, the only non-vanishing are
\begin{equation}
R_{\; aab}^{a}=-G^{-1}G_{,ab}+G^{-2}G_{,a}G_{,b},
\label{(B12)}
\end{equation}
\begin{equation}
R_{\; \Theta \varphi \Theta}^{\varphi}=1-2G^{-1},
\label{(B13)}
\end{equation}
\begin{equation}
R_{\; \varphi \Theta \varphi}^{\Theta}
=(1-2G^{-1})\sin^{2} \Theta,
\label{(B14)}
\end{equation}
and hence
\begin{equation}
R=-2G^{-2}\left(G_{,ab}-{G_{,a}G_{,b}\over G} \right)
+2(a+b)^{-2}\left(1-{2\over G}\right).
\label{(B15)}
\end{equation}
With the notation in Eqs. (3.14)--(3.16), we find
\begin{equation}
G_{,a}=G \left[{2\over (a+b)}+{g_{1,a}\over g_{1}}
+{F_{,a}\over F} \right],
\label{(B16)}
\end{equation}
\begin{equation}
G_{,b}=G \left[{2\over (a+b)}
+{g_{2,b}\over g_{2}}+{F_{,b}\over F} \right],
\label{(B17)}
\end{equation}
\begin{equation}
G_{,ab}= {G_{,a}G_{,b}\over G}+G \biggr[-{2\over (a+b)^{2}}
+{F_{,ab}\over F}-{F_{,a}F_{,b}\over F^{2}} \biggr].
\label{(B18)}
\end{equation}
Therefore Eq. (B15) yields
\begin{equation}
R(\theta)= (a+b)^{-2}\biggr[2-{1\over g_{1}g_{2}F^{2}} \biggr(
F_{,ab}-{F_{,a}F_{,b}\over F}\biggr) \biggr].
\label{(B19)}
\end{equation}
For example, in general relativity  
\begin{equation}
F_{,a}={\partial F \over \partial f}{\partial f \over \partial w_{0}(f)}
{\partial w_{0}(f)\over \partial w_{0}(a)}
{\partial w_{0}(a)\over \partial a}
=g_{1}(a)(2f-6Mf^{2})F
\label{(B20)}
\end{equation}
by virtue of 
\begin{equation}
{\partial w_{0}(f) \over \partial f}
=-{1\over f^{2}}+{1\over \left(f^{2}-{f\over 2M}\right)}=-{1\over F}, \;
{\partial w_{0}(a)\over \partial a}=-g_{1}(a),
\label{(B21)}
\end{equation}
and hence
\begin{equation}
F_{,ab}-{F_{,a}F_{,b}\over F}=g_{1}(a)(2f-6Mf^{2})_{,b}F
=g_{1}(a)g_{2}(b)(2-12Mf)F^{2},
\label{(B22)}
\end{equation}
\begin{equation}
R(0)={12 Mf \over (a+b)^{2}}.
\label{(B23)}
\end{equation}
Our definition of scalar curvature has opposite sign with respect to the
work in \cite{Stew79}, but of course this does not affect the results.

In section 3, Eq. (B19) yields instead
\begin{equation}
R(\theta)={12mf \over (a+b)^{2}},
\label{(B24)}
\end{equation}
where $m$ is the mass function defined in Eqs. (1.6) and (3.14).

\end{document}